 \documentclass[final,5p,times,twocolumn]{elsarticle}

\usepackage{amssymb}
\usepackage{units}
\usepackage{hyperref}

\begin{document}

\begin{frontmatter}



\title{Summary of Working Group 8: Advanced and Novel Accelerators for
High Energy Physics}

\author[LPGP]{B. Cros}
\author[MPP,CERN]{P. Muggli}
\author[LBNL]{C. B. Schroeder}
\author[Tsinghua]{C. Tang}

\address[LPGP]{LPGP CNRS, Universit{\'e} Paris-Sud, Universit{\'e} Paris Saclay, Orsay, France} 

\address[MPP]{Max Planck Institute for Physics, M{\"u}nchen, Germany}

\address[CERN]{CERN, Geneva, Switzerland} 

\address[LBNL]{Lawrence Berkeley National Laboratory, Berkeley, California, USA} 

\address[Tsinghua]{Tsinghua University, Beijing China}

\begin{abstract}
We briefly summarize the work and discussions that occurred during the
Working Group 8 sessions of the EAAC 2017, dedicated to advanced and
novel accelerators for high energy physics applications.
\end{abstract}


\end{frontmatter}


\section{Introduction}
\label{sec:intro}
Working Group 8 (WG8), Advanced and Novel Accelerators for High Energy
Physics, was created as a result of the Advanced and Novel
Accelerators for High Energy Physics Roadmap (ANAR) Workshop 2017 that
took place at CERN, April 25-28, 2017~\cite{anar}. The workshop was
organized at the initiative of the ICFA subpanel on Advanced and Novel
Accelerators chaired by B.\ Cros. The workshop was a first step
towards federating the fragmented advanced and novel accelerator (ANA)
community, to define an international roadmap towards colliders based
on advanced accelerator concepts, including intermediate milestones,
and to discuss the needs for international coordination. A preliminary
report summarizing the state-of-the-art of ANAs and the proposed
future steps can be found at~\cite{anardoc}.

Working Group 8 was organized in two specific parallel sessions and
one joint session with Working Group 1 - Electron beams from plasmas.
The WG8 charge was to examine key challenges, discuss suitable
concepts, and identify topics for future R\&D or innovation, including
electron and positron sources, damping rings, optics between stages,
acceleration of positrons, luminosity goals, final focus, and overall
efficiency.

\section{General Discussion}
\label{sec:gen}
The work started with a general discussion led by A. Seryi who
emphasized the need for our ANA community to take advantage of the
vast knowledge of the conventional accelerators community. That
community has built working accelerators, but has also developed
Conceptual Design Reports (CDRs) and Technical Design Reports (TDRs)
for future colliders that are ready to be built, such as the
International Linear Collider (ILC) and the Compact Linear Collider
(CLIC).

In the ILC case, the enterprise started with a group of motivated
scientists who produced more and more refined designs that were
periodically and critically reviewed by the community over many years.
These designs were organized around systems (injector, damping rings,
linac, beam dynamics, beam delivery system, etc.). Each system was
analyzed, and for each system issues were identified and ranked.
Required R\&D programs relevant to collider criteria such as
feasibility, reliability, production, and technical and cost
optimization were defined and carried out. The work started with a
number of options for each system, e.g., warm and superconducting
technology for the linac. From the many designs that were generated
(Tesla, JLC, NLC, etc.), one was eventually selected to become the
ILC.

The ILC process toward a collider could serve as model for the design
of an ANA-based advanced linear collider (ALC). Indeed, at this time
the community includes four major ANA R\&D efforts: Laser-driven plasma
wakefield accelerators (LWFA), particle-driven plasma wakefield
accelerators (PWFA), structure wakefield accelerators (SWFA), and
dielectric laser accelerators (DLA), all with potentials and
challenges, as summarized, for example, in the ANAR
report~\cite{anardoc}.

This discussion helped define the structure of the next workshop
organized by ICFA-ANA and the ALEGRO (Advanced LinEar collider study
GROup) group~\cite{alegro}, to be hosted by the University of Oxford
and the John Adams Institute March 26-29, 2018. The upcoming ALEGRO
workshop will have seven main working groups: Physics Case (PC),
Collider machine design/definitions (CMD), Theory, Modeling,
Simulations (TMS), LWFA, PWFA, SWFA, DLA, and a joint sub-WG on
positron acceleration (PAC).

\section{Plasma-based ANAs}

In this section, we report on the conclusions of the presentations
discussing acceleration in plasma.

In plasma-based ANAs operating in the bubble regime, the transverse
fields are of the same order as longitudinal fields. Whereas it can be
seen as an advantage to channel bunches over long distances, it also
means that the particles emit betatron radiation. The radiated power
depends on the particle energy and amplitude of oscillation about the
beam axis. Calculations performed by V.\ Shpakov show that, per unit
length, the radiation loss is always smaller than the energy gain and
betatron radiation loss is therefore not a limitation to energy gain.
As expected, emission betatron radiation together with acceleration
lead to some cooling of the bunch emittance. However, radiation loss
causes increase of the energy spread, which decreases the overall
quality of the accelerated bunch.

Under-dense plasma lenses have very strong focusing gradients compared
to magnetic lenses. The focusing gradient, proportional to the local
plasma electron density, can in principle be tailored along the lens.
Increasing the focusing gradient \emph{adiabatically} along the length
allows for minimization of synchrotron radiation and overcoming of the
Oide limit. Calculations, presented by F.\ Filippi, show that for an
experiment at the INFN SPARC-LAB, an exponential increase of the
plasma density by a factor of 400 over 6\,cm and starting with a
density of 5$\times$10$^{15}$\,cm$^{-3}$ would be needed to focus a
110\,MeV electron bunch from an initial 75\,$\mu$m radius to $\sim
5\mu$m. Using a capillary discharge with a tapered entrance tube could
be used for the density tapering. The size of the focused bunch could
be inferred from the spectrum of the radiation emitted by the
electrons in the constant density plasma inside the capillary. 

Nikolay Andreev discussed external injection and acceleration of short
electron bunches in plasma waves generated by intense femtosecond
lasers in plasma channels. The dependence of the bunch length and
energy spread was studied as a function of the phase of the plasma
wave in the quasi-linear regime where the electrons are injected. For
the parameters studied, when an electron bunch is injected with a size
of the order of 20\% the plasma wave length, at the maximum of the
wave potential, the energy gain is or the order of 2~GeV over 20~cm of
propagation with an rms energy spread smaller than $0.01$. This energy
spread is achieved for charges of the order of a few pC; for larger
charges, beam loading leads to increased energy spread. Mechanisms
relevant to multi-stage accelerators, such as the preservation of
emittance and spin polarization of electron beams between plasma
stages were examined by D.\ Pugacheva. Multi-stage LWFA was shown to
preserve beam polarization, as the theoretical prediction gives 0.02\%
for depolarization after acceleration up to 1 TeV, which is suitable
for high-energy-physics applications.

Driving plasma wakefields with a positively charged particle bunch
raises a number of challenges related to the nonlinearity of the
transverse forces acting on the bunch when reaching the non-linear
regime. Some of the challenges can be overcome by using a hollow
plasma channel rather than a uniform plasma. High-energy proton
bunches are long, which means that one has to rely on self-modulation
to produce a train of bunches that can then drive large amplitude
wakefields. However, the self-modulation process operates by
de-focusing a large fraction of the protons. Y.\ Li proposed to deal
with these two issues by driving wakefields with a pre-formed train of
short proton bunches in a hollow plasma channel. Initial estimates
also show that this scheme is relatively weakly sensitive to
variations of the plasma channel.

\section{DLA}

The dielectric laser accelerator (DLA) scheme has shown accelerating
gradients exceeding GV/m, however over rather short accelerating
distances and with small energy gains (sub-MeV). One of the advantages
of the DLA is that optical elements can be used for all the functions
necessary for an accelerator (power source, focusing, diagnostics,
etc.), potentially allowing for an "on-chip" accelerator. As with
other ANA concepts, one of the next steps for DLA is a global concept
that includes an injector and a number of acceleration stages to
produce high-energy, high-quality bunches. The DLA community, gathered
in the ACHIP collaboration, is addressing challenges related to
developing a high-energy accelerator on a chip.

J.\ England summarized the ACHIP program, including issues such as
acceleration from rest to relativistic energies, micro-bunching at the
laser wavelength scale, powering of multiple structures, compensation
of globally de-focusing transverse forces during the
weakly-relativistic part of the acceleration process, etc. The results
of these studies are continuously integrated into updated versions
of the "straw-man" concept collider based on DLA.

\section{Simulations}

A key component for the progress towards an Advanced Linear Collider
based on one of the ANAs is the availability of computationally
efficient simulation tools. Simulations are particularly challenging
for plasma-based ANAs.

Most simulation codes use explicit particle-in-cell (PIC) algorithms
that are particularly consumming in terms of computer memory and time.
Reduced models (quasi-static, hybrid, etc.) and numerical methods
(boosted frame, azimuthal decomposition, etc.) can provide significant
time and memory savings, without loss of physical fidelity. However,
at present, the full modeling of a single multi-GeV accelerator stage
in 3D remains extremely challenging. J.-L.\ Vay presented a very
ambitious program, a seven-year, exascale modeling project, based on
WarpX. The goal of the project aims at a full numerical description
(3D) of 100 plasma-based accelerator stages, toward a 1~TeV collider
design. The project plan includes modeling of a single plasma
accelerator stage with static mesh refinement by 2017, modeling a
single GeV-scale plasma accelerator stage by 2018, and modeling of
multiple GeV-scale plasma accelerator stages by 2020. WarpX is an
up-graded version of the WARP, coupled to the latest adaptive mesh
refinement algorthms. WARP is an open-access code, and WarpX is
scheduled to be released to the public by 2018. The plans also include
visualization tools as well as interfaces for analysis with various
programs. This is a very important and welcome initiative.

\section{Facilities}

During the joint session with Working Group 1 (Electron beams from
plasmas) of some of the present and planned ANA experimental
facilities were described.

EuPRAXIA is a design study for a 5~GeV electron beam research facility
for light source (FEL) and high-energy physics applications. The major
goal for EuPRAXIA is to design and, in a second phase, build a compact
European advanced accelerator with superior beam quality. P.\ A.\
Walker presented a designed and preliminary global parameters of a
generic facility that uses ANA technology to drive an FEL. The
facility would generate 5~GeV electron beams and include two users
areas: one for the FEL and one for high-energy physics applications.
The design accommodates options for the electron source (RF-gun with
x-band linac, LWFA), and for the accelerator (LWFA, PWFA). Drive
lasers and klystrons would be located in an upstairs mezzanine,
whereas accelerators, undulators and experimental areas would be
located on the ground floor.

At the moment there are five European facilities that could candidate
to host the construction of a facility resulting from Eupraxia design
study.

M. Ferrario presented plans for a new compact FEL facility at INFN-LNF
that would be compatible with the EuPRAXIA design. The plans include
all the options envisaged by EuPRAXIA, but also the option to have a
1~GeV x-band linac that could drive the FEL without plasma
acceleration. The TDR is expected to be completed by the end of 2017.
The facility would be an upgrade of the current SPARC-LAB that already
operates with an s- and c-band accelerator to drive an FEL. In
addition, the high-power laser FLAME, presently used for LWFA studies,
would be upgraded with the aim of externally injecting and
accelerating the bunch from the RF-gun. Plasma sources are already
available for PWFA and plasma lens studies. Preliminary simulation
results indicate that beams produced in the current SPARC-LAB, and
thus its upgraded version, from plasma accelerators would lead to
lasing of the FEL in the water window.

FLASHForward is the future-oriented wakefield accelerator research and
development facility at FLASH, DESY. The main goal of the project is
to explore PWFA. At FLASHForward, a PWFA witness bunch can be
generated by a plasma cathode or from the FLASH linac. Single bunches
can in principle be tailored for large transformer ratio and optimal
beam loading experiments. Studies will also focus on the hosing
instability, a effect that could disrupt beams in very long
acceleration sections. In a second phase (starting 2020), undulators
will be added to the beamline to drive an drive an FEL using an
electron bunch produced in the PWFA. The work is supported by a strong
simulation program, including start-to-end simulations. In his
presentation J.\ Osterhoff described how many of the critical
scientific challenges faced by the PWFA, and other ANAs, can be
addressed at FLASHForward.

ARES is another facility at DESY studying LWFA and possibly DLA for
FELs and other applications. The beam produced by the ARES linac would
be injected into the LWFA. Towards this end, simulations including
deleterious effects such as CSR, presented by J.\ Zhu, showed that
after compression the ARES bunch could reach sub-fs length, which
would make the bunch suitable for injection into an LWFA.
These short bunches contained 0.8~pC, with 600~A current. 
Longer bunches with larger currents can also be produced and, perhaps,
are more suitable to drive the FEL. Maintaining a low energy spread
suitable for FEL lasing is a challenge, as in all LWFA designs. In the
ARES case, simulations showed space charge effectscan limit the bunch
current and the beam matching into the plasma.

The MAX IV is a 3~GeV light source storage ring located in Lund
(Sweden). The 3~GeV beam from the linac is available for ANA
experiments, when not used for filling of the ring. Simulation results
presented by O.\ Lundh show that the 100~pC bunch can be compressed to
reach more than 10~kA of peak current. However, in this case emittance
growth becomes a serious issue. Less aggressive compression to 1.5~kA
produces a bunch with low emittance and relative energy spread.
Driving the gun photo-cathode with two laser pulses could lead, after
two stages of compression to a drive/witness bunch train that would be
very interesting for PWFA experiments. In particular, simulations
showed that a loaded accelerating field could reach the 10~GV/m level.
Plans for PWFA experiments are awaiting endorsement by the MAX IV
Laboratory.
 
\section{Conclusions}
\label{sec:conclusions}

Discussions during the working group re-enforced the need for active
collaborations, community-gathering behind a global high-energy
physics application of ANAs, continuous critical analysis of the
various "straw-man" designs for an ALC, and need for leadership in
this endeavor. It is clear that there is much to learn from the
conventional accelerator community's experiences builing accelerators
and developing TDRs for future linear colliders (e.g., CLIC and ILC).
It is also clear that there is an urgent need to involve this
community in the development of ANAs and also in the development of a
global collider concept starting at the injector and ending at the
detector and physics case. This model will be used for the up-coming
ALEGRO workshop and for the generation of a document as an input for
the next European Strategy on High-Energy Physics.

From the scientific point of view, presentations in WG8 showed that
significant progress is being made towards high-energy accelerator
concepts. The most attractive feature of ANAs remains the large
accelerating gradients.

There are now roadmaps for ANA development that have been drawn in the
US \cite{us-roadmap} and in Europe \cite{anardoc}. There is a European
facility design study (EuPRAXIA) focusing on a plasma-based ANA to
produce high-quality beams.
A number of facilities are either operating, coming on-line, in the 
building phase, or in the planing phase. %
All of these facilities are geared toward light sources or high-energy
physics applications. They will all contribute to addressing the major
scientific challenges that have been identified by the community, most
recently at the ANAR Workshop. ALEGRO will provide a forum for
organizing the global ANA efforts towards producing high-quality and
high energy beams, as well as deliver a document for the next European
Strategy input and for an ALC.

\appendix


\section{References}


\begin{thebibliography}{00}
\bibitem{anar}\url{https://indico.cern.ch/event/569406/}
\bibitem{anardoc}\url{http://www.lpgp.u-psud.fr/icfaana/ANAR2017_report.pdf}
\bibitem{alegro}\url{https://indico.cern.ch/event/677640/}
\bibitem{us-roadmap}\url{https://science.energy.gov/~/media/hep/pdf/accelerator-rd-stewardship/Advanced_Accelerator_Development_Strategy_Report.pdf}
\end{thebibliography}
\end{document}